\documentclass[epj]{svjour}
\usepackage{graphicx}
\usepackage{dirtytalk}
\usepackage{amsmath,amssymb}
\usepackage{hyperref}
\usepackage{xcolor}

\newcommand{\cred}[1]{{#1}}

\newcommand{\neanb}{$^{22}$Ne($\alpha$,n)$^{25}$Mg }
\newcommand{\ctanb}{$^{13}$C($\alpha$,n)$^{16}$O }
\newcommand{\ctan}{$^{13}$C($\alpha$,n)$^{16}$O}
\newcommand{\iso}[2]{$^{#1}$#2}

\begin{document}

\title{s-process nucleosynthesis in low-mass $AGB$ stars by the $^{13}$C($\alpha$,n)$^{16}$O  neutron source.}
\author{Inma Dom\'{\i}nguez\inst{1} \and Carlos Abia\inst{1} \and Maurizio Busso\inst{2,3} \and Sara Palmerini\inst{2,3,4} \and Oscar Straniero\inst{5,6}} 
{}                     
%
%
\institute{Dpt. F\'\i sica Te\'orica y del Cosmos. University of Granada, 18071 Granada, Spain \and Dpt. of Physics and Geology, University of Perugia, via A. Pascoli s/n, 06125 Perugia Italy 
\and INFN Section of Perugia,  via A. Pascoli s/n, 06125 Perugia Italy 
\and INAF, Osservatorio Astronomico di Roma, via Frascati 33, 00078 Monte Porzio Catone (Rome), Italy 
\and INAF, Osservatorio Astronomico d'Abruzzo, Via Mentore Maggini, 64100, Teramo, Italy 
\and INFN, Sezione di Roma, Piazzale A. Moro 2, 00185 Roma, Italy
}
\date{Received: date / Revised version: date}
%
\abstract{In this review we outline the temporal growth of our knowledge on \textit{slow neutron captures} (the so-called s-process), concentrating on its main part occurring during the final stages of stellar evolution for low or intermediate-mass stars when they approach for the second time the Red Giant Branch and are therefore called \textit{Asymptotic Giant Branch}, or AGB, stars. In particular, we focus our attention on how, in this field,  studies passed from a first era of inquiries based on nuclear systematics (now often referred to as the \textit{the phenomenological approach}), to numerical nucleosynthesis computations performed in stellar codes. We then discuss how these last were forced, by observational constraints, to almost abandon, for the synthesis of nuclei between Sr and Pb (i.e. the $main$ $component$), the rather naturally activated \neanb neutron source (operating efficiently at $T$ ${\gtrsim}$  3.5 ${\cdot}$ 10$^8$ K, i.e. 30 keV, and producing a neutron density $n_n \gtrsim 5 \cdot 10^8$ cm$^{-3}$). This implied considering the alternative reaction \ctan, that can be activated locally after each of the recurring mixing episodes from the envelope (collectively referred to as the {\it Third Dredge Up}, or $TDU$). The mentioned crucial reaction occurs at a relatively low temperature ($T \simeq 8-9 \cdot 10^7$ K., i.e., less than 8 keV), in the time intervals separating two subsequent $thermal$ $instabilities$ of the He shell (also named $Thermal$ $Pulses$, or $TP$). The layers where \ctanb  operates are characterized by a radiative equilibrium and their low temperature also yields low values for the neutron density ($n_n \lesssim 10^7$ cm$^{-3}$). The activation of such a neutron source is unfortunately  not straightforward, as little $^{13}$C is left behind by shell H burning in the zone bracketed by the two alternatively burning shells of AGB stars. One has therefore to discover the proper mixing mechanisms providing further proton captures on the abundant $^{12}$C there present. Despite this difficulty, the modelling of s-process as provided by this alternative neutron-producing reaction was crucial to clarify the origin and the distribution of nuclei from Sr up to Pb and Bi in the Galaxy, hence we outline the various (mainly non-convective) mixing processes so far considered for this purpose and their relative efficiency. We conclude by accounting for the extensive observations and measurements on several sources, from low-metallicity stellar objects to presolar grains, from normal AGB stars to post-AGB sources and binary systems, which motivated the crucial change of paradigm from the \neanb to the \ctanb neutron source.
}
\PACS{Gravitation, Cosmology and Astrophysics: Nuclear Astrophysics: Hydrostatic stellar nucleosynthesis: s-process}
%
\authorrunning{Dom\'{\i}nguez, I. et al.}
\titlerunning{Production of n-rich nuclei in AGB stars}
\maketitle
\section{Introduction}
\label{intro}
For the most abundant nuclei in the solar composition production derives from fusion reactions, generally also releasing the energy required to power stellar evolution. Early studies of such processes marked the beginning of what is now called {\it nuclear astrophysics}. It all started from the invention of the mass spectrometer \cite{ast} by the British chemist Francis William Aston (Nobel Prize for Chemistry in 1922), through which he discovered that a helium nucleus has a mass lower than that of four separated protons. This fact induced Eddington \cite{Eddin} and Perrin \cite{Perrin}, almost contemporaneously, to advance the suggestion that hydrogen fusion in the Sun might produce the energy accounting for its luminosity, by releasing the mass excess in the form of radiation. Subsequent works by \cite{gam28} and \cite{ah29} on the Coulomb barrier to be overcome and on the required quantum {\it tunnel effect} were instrumental in the first suggestions on the mechanisms of nuclear fusion at solar temperatures by \cite{a36}, \cite{bc38} and \cite{w39}, \cite{b39}. These early studies were the basis on which Fred Hoyle built his theory of stellar nucleosynthesis \cite{hoy46}, \cite{hoy54}.

However, elements more massive than those in the iron group
have a binding energy that decreases with increasing mass. This prevents the synthesis of the heaviest nuclides from occurring through the fusion reactions typically active among lighter species in stellar interiors, up to the demise of the parent star. In fact, for the most massive isotopes fusion becomes endothermic; and in any case, it would also require extremely high temperatures to overcome their enormous Coulomb barrier.  Specifically, for a long time, the most tightly bound nuclide was believed to be $^{56}$Fe itself, until \cite{few} showed that the crucial nucleus was the nearby $^{62}$Ni, the second being $^{58}$Fe (which is also stable to $\beta$- decays). In any case, the binding energy decline from Ni on requires processes different than fusion to occur in stars. In the seminal work by \cite{b2fh} (hereafter B$^{2}$FH, from the authors' initials) such very heavy nuclei from Fe to actinides were shown to be mainly synthesized by neutron capture processes. These latter are broadly divided into two categories, respectively named {\it fast} and {\it slow}, depending on the relative average speed of neutron captures and ${\beta}$ decays along their formation paths. Donald Clayton  et al. performed a first detailed analysis of these processes at the beginning of the 1960s \cite{CFHZ}, even before Clayton himself achieved his PhD on the same subject \cite{DC_thesis}. That fundamental theoretical analysis remains famous today as the CFHZ paper (a name again derived from the authors' initials).


That work introduced a solution to the simplified s-processing equation (see Section \ref{sec2}), based on an exponential distribution of neutron exposures. It was then followed by various studies on neutron captures based on the same idea, confirming and specifying the earlier findings, up to the synthesis published by \cite{seeg65}, containing a full account of neutron capture processes. This work presented an analysis of the solar $\sigma N_s$ curve, fitting it by an exponential solution for the "s-only" nuclei (those exclusively due to the s-process), showing that the theoretical curve also fitted the s-component of the other nuclei in the solar composition \cred{(see Fig. 1 in \cite{seeg65}). This fundamental result is exhaustively illustrated also in Clayton's famous book \cite{clay68}, which remains even today a crucial reference on these topics.}

Seeger et al. \cite{seeg65}, studying the path through Sr isotopes or the alternative through Rb isotopes, also introduced a first discussion of {\it reaction branchings}, i.e. of the points in the s-process path where $\beta$-decays or other nuclear reactions become of comparable speed with (n,$\gamma$) captures, so that the nuclosynthesis flux splits into different channels.

The last years of the 1960s saw the first quantitative calculations made to describe how s-processing could occur in stars. The final evolutionary stages of red giants have been observed to be a site where slow neutron captures had to occur due to the presence on the surface of the unstable element Tc \cite{mer}. Then the first attempts to model their rather complex evolution by \cite{SH} showed their structure, namely their double-shell sources (burning H and He, respectively) to be prone to recurrent instabilities (often called {\it thermal pulses} or $TP$), hence the name of this evolutionary stage as {\it Thermally Pulsing Asymptotic Giant Branch, or $TP-AGB$}  phase. See later, section \ref{sec3} for a discussion of these phenomena). 

Let us now have a closer look, in the next sections, at the critical steps that characterise the passage through the phenomenological treatment and the early models for AGB star evolution. We shall again follow these issues historically, starting from what had been deduced by nuclear systematics.

\section{The phenomenological approach}
\label{sec2}

As mentioned, the first outline of a mathematical formulation for the s-process was due to CFHZ and \cite{seeg65} and is now generally called the {\it phenomenological approach} to neutron captures. The analysis by the above authors starts from the consideration that, in the simplest possible representation of neutron captures occurring on two adjacent stable nuclei of atomic mass $A$ and $A-1$, the abundance by number 
$N(A)$ of the nucleus $A$ varies in time as:

\begin{equation}
\frac{dN(A)}{dt} = N(A-1) n_n <\sigma(A-1)\cdot v> - N(A) n_n <\sigma(A)\cdot v> \label{eq1}
\end{equation}

\noindent
where $<\sigma(A) \cdot v >$ is the Maxwellian-average product between the neutro$n$-capture cross section for the nucleus $A$ and the plasma velocity, while $n_n$ is the neutron density. One can then define the {\it neutron exposure} $\tau$, i.e. the time integrated neutron flux, to be:

\begin{equation}
\tau = \int_0^t{n_n v_T dt}
\end{equation}

\noindent
where $v_T$ indicates the thermal velocity. Subsequently, by introducing the $average$ cross section, $\hat{\sigma}(A)$:

\begin{equation}
\hat{\sigma}(A) = \frac{< \sigma \cdot v>}{v_T}
\end{equation}
\noindent
equation (\ref{eq1}) becomes:

\begin{equation}
\frac{dN(A)}{d\tau} = N(A-1) \hat{\sigma}(A-1) - N(A) \hat{\sigma}(A) \label{eq2}
\end{equation}
When the neutron flow is  in steady-state conditions (i.e. $dN/d \tau = 0$), equation (\ref{eq2}) yields:
\begin{equation}
\hat{\sigma}(A) N(A) = const.
\end{equation}
\noindent
which tells how, if slow neutron captures occur at equilibrium, then we expect that the resulting solar $\sigma N$ products of stable nuclei affected only by neutron captures remain constant. 

The above condition is actually met (roughly) experimentally, for isotopes that lay far from the regions
where the shell model foresees the closure of neutron shells, in which jumps in the nuclear-charge radius occur and very stable nuclear structures are found. This last situation characterizes nuclei that contain specific numbers of neutrons, namely $N = 2, 8, 20, 28, 50, 82, 126$ (called {\it magic numbers}). 

\cred{These jumps are, for example, evident for $N=50, 82$, and $126$, in Fig. 1 of \cite{seeg65}. They are even clearer in the similar plot reported by \cite{ka1} in their Fig. 3, where reaction branchings are properly accounted for.} The jumps actually correspond to nuclei that reveal their high stability through relatively large abundances and very small cross sections (few millibarns). More generally, the same authors \cite{seeg65} showed how the experimental solar products $\sigma N$ could be mimicked by any distribution of neutron exposures decreasing for increasing values of $\tau$, like e.g. in a power-law with a negative exponent or in a negative exponential form. In particular, they suggested to approximate the experimental data through the curve:

\begin{equation}
\rho(\tau) = \frac{GN^{56}_{\odot}}{\tau_0} exp(-\tau/\tau_0) \label{eq6}
\end{equation}

\noindent Here, a fraction $G$ of the solar $^{56}$Fe abundance, $N^{56}_{\odot}$ (the main nucleus on which $n$-captures occur)  is subject, in the s-process site, to an exponential series of neutron irradiations, and $\tau_0$ is a free parameter (called {\it mean neutron exposure}). Subsequently, \cite{cw74} found that, adopting expositions as in equation (\ref{eq6}), the $\sigma N_s$ products ($N_s$ being the fractional abundance of the various s-process isotopes) could be analytically expressed using the formula: 
\begin{equation}
\sigma N_s = \frac{GN^{56}_{\odot}}{\tau_0} \prod_{i=56}^{A}\left( 1 + \frac{1}{\sigma_i \tau_0} \right)^{-1}
\end{equation}
It became also clear that, under the above conditions, more than one astrophysical mechanism would be required to overcome the  \say{bottlenecks} represented by magic nuclei, and those authors suggested 
a superposition of three exponential {\it components}, namely a  \say{weak}  one, for nuclei up to Sr (with a number of neutrons $N \lesssim 50$), a  \say{main} one, for nuclei above Sr and lighter than Pb (with $50 \lesssim N \lesssim 126$) and a  \say{strong}  one, mainly required for the magic $^{208}$Pb.


After the first anticipations by \cite{seeg65}, already quoted, a full treatment for branching points, where the chain of neutron captures meets $\beta$-unstable isotopes and a competition between capture and decay becomes possible, was then considered by \cite{wn78}, thus opening the possibility to estimate the environmental conditions (neutron density and temperature) in the original stellar sites, from estimates of the {\it branching ratio} $f_{-}$:
\begin{equation}
f_{-} = \frac{\lambda^-}{(\lambda^- + \lambda_n)}
\end{equation}
\noindent
where $\lambda^{-} = 1/\tau_{\beta^{-}}$ is the rate of beta decays and $\lambda_n = n_n <\sigma v >$ is the rate of neutron captures.

From the above purely nuclear considerations, and remembering that many $\beta$-decay rates depend on the temperature $T$, four characterizing parameters ($G, \tau_0, n_n$ and $T$) could be inferred, thus providing tools useful to guide astrophysicists in their search for suitable stellar models. 

\section{Early nucleosynthesis models for AGB stars}
\label{sec3}
Soon after the hypothesis of exponential distributions of exposures was advanced by \cite{seeg65},  stellar models were extended to the final evolutionary stages of low and intermediate-mass stars, noting the \say{peculiar} behavior of the AGB stages powered by nuclear burning of H and He, in a double-shell structure.

\cite{SH} underlined a peculiarity affecting strongly the s-process, called {\it thin shell instability}.
It occurs when, in  spherical  symmetry, a constant mass $\Delta m$ is concentrated at a constant radius $r$ in a shell of thickness $l << r$ ($\Delta m \simeq 4 \pi \rho r^2 l$) and is in hydrostatic equilibrium, so that:
\begin{equation}
\frac{dP}{P} = -4\frac{dr}{r}
\end{equation}
Then one has:
\begin{equation}
\frac{d\rho}{\rho} = -\frac{dl}{l} \simeq -\frac{dr}{l} = -\frac{r  dr}{l  r}
\end{equation}
Hence:
\begin{equation}
\frac{dP}{P} \simeq 4 \frac{l}{r }\frac{d\rho}{\rho}
\end{equation}
In the stellar plasma one also has:
\begin{equation}
\frac{dP}{P} = \alpha\frac{d \rho}{\rho} + \beta\frac{dT}{T}
\end{equation}
Hence:
\begin{equation}
\left(4\frac{l}{r}-\alpha\right)\frac{d\rho}{\rho} = \beta \frac{dT}{T}
\end{equation}
For thermal stability one would require 
\begin{equation}
4\frac{l}{r} > \alpha
\end{equation}
\noindent
which, for $l/r \rightarrow 0$, is no longer satisfied. This drives an unstable situation. Suppose indeed that the layer expands. If the pressure from hydrostatic equilibrium drops with radius faster than that due to the expansion, then the layer will continue to expand
and cool out of equilibrium. If not, then the temperature will rise, inducing a thermal instability and an ensuing sudden enhancement of the luminosity, because of a {\it thermonuclear runaway}, before relaxation occurs reestablishing the previous condition. See e.g. \cite{str23} for a discussion.

\begin{figure}
    \centering
    \resizebox{\columnwidth}{!}{\includegraphics{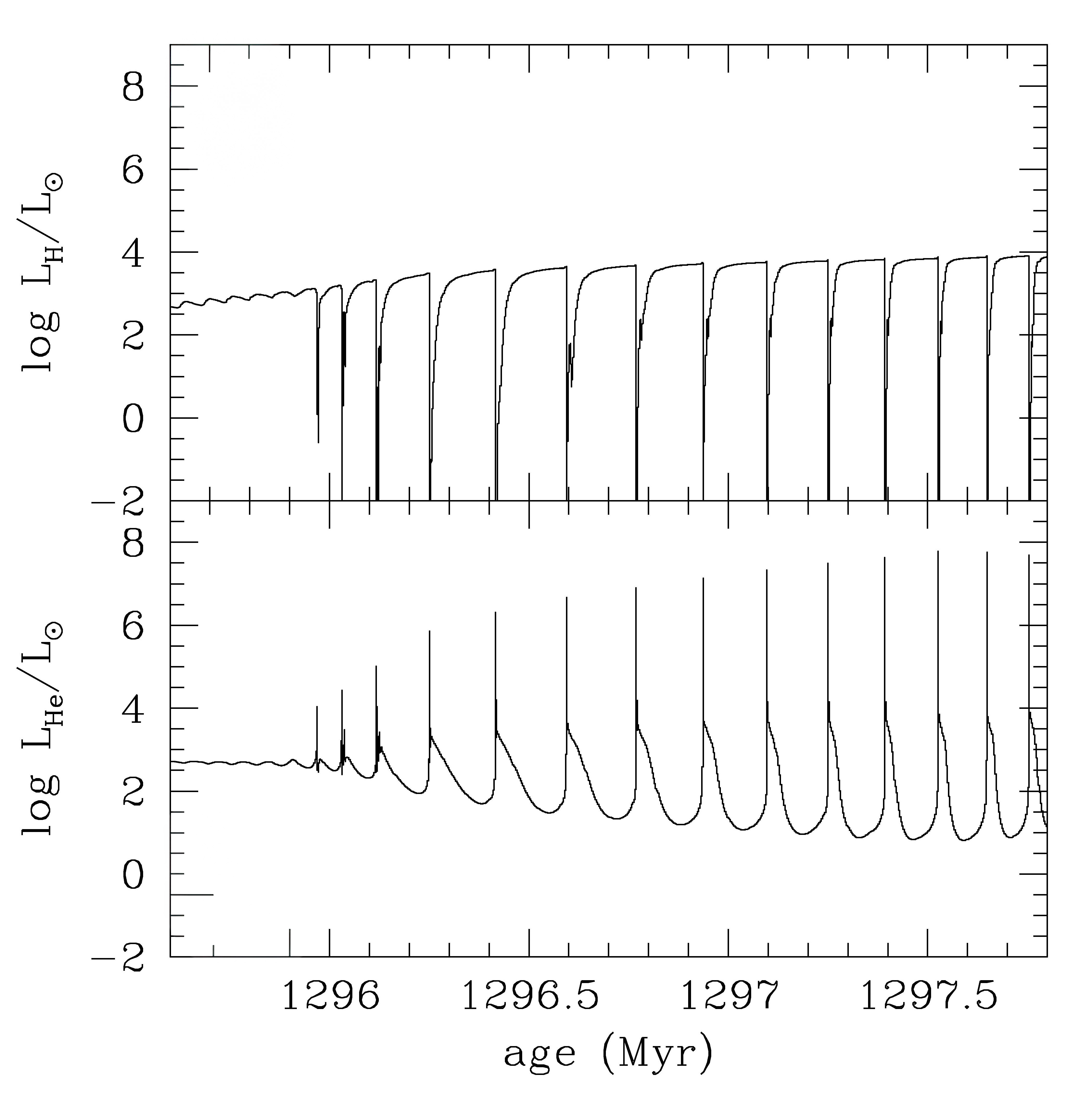}}
\caption{Lower panel: the recurrent occurrence of thermonuclear instabilities in the He-shell luminosity of a low-mass AGB star ($M=2$ M$_\odot$). Upper panel: the corresponding variation of the H-shell luminosity. Note the sudden increase of the nuclear energy released by the He burning (up to $10^8$ L$_\odot$), causing the expansion of the more external layer that becomes progressively cooler, until the H-burning shell dies down. At that time, the convective envelope penetrates inward, moving the ashes of the He burning upward (see Fig. \ref{fig4})}
\label{fig3}
\end{figure}

\begin{figure}
\includegraphics[width=\linewidth]{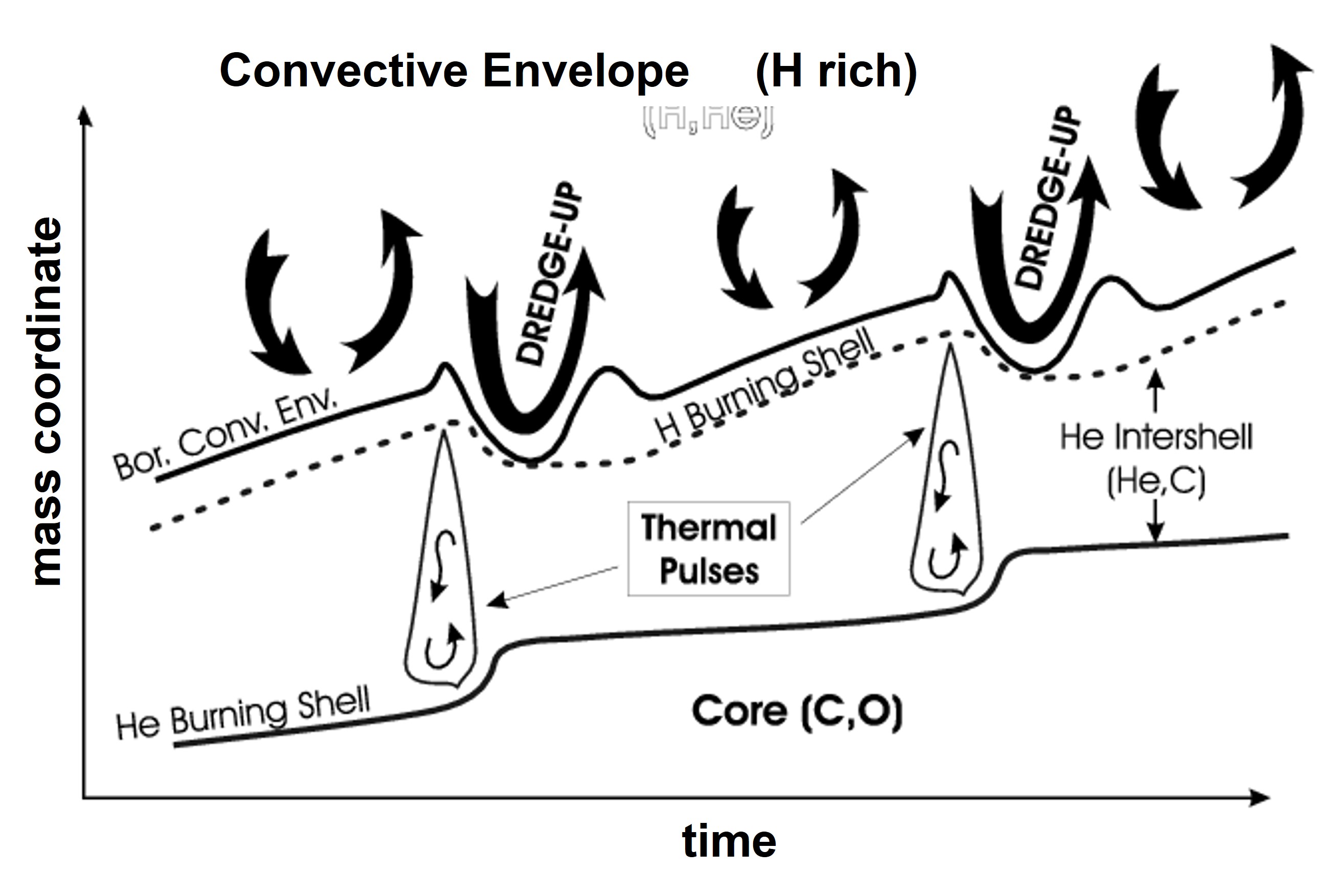}
\caption{A sketch of two successive thermal instabilities (thermal pulses), developing convective zones that mix the whole He-rich mantel. Note: a) that the second convective zone partially overlaps the region mixed by the previous one and b) the inward penetration of the convective envelope after each thermal pulse (the {\it third dredge-up}).} 
\label{fig4}
\end{figure}

The repeated occurrence of this unstable condition during the evolution of the star is shown in Fig. \ref{fig3}. As a consequence of this peculiarity, low and intermediate-mass stars in their final stages of evolution (the TP-AGB stages) are characterised by recurrent sudden luminosity flashes, accompanied by the development of intermediate convective zones ($ICZ$), each followed by the penetration of the convective envelope below the upper border of the He-rich layers (see Fig. \ref{fig4}).

As the figure shows, He-burning (which is normally inhibited by the core degeneracy) restarts abruptly (in a semi-explosive way) 
after long periods of H-burning activity, which compresses and gradually heats the He-rich layers. Hence, He re-ignition drives the mentioned instability, in which a peak in temperature occurs (starting from about 1.5 10$^8$ K before the instability and reaching up to $2.8 - 3.5$ $10^8$ K at maximum efficiency, mainly depending on the stellar mass). 

The large energy output thus generated (see also Fig. \ref{fig3}) induces the star to develop an $ICZ$ to increase the efficiency of the energy transport, so that the products of He-burning are spread throughout the intershell layers. These last then expand and cool, and the H-burning shell is switched off, while the convective envelope penetrates below the H-He discontinuity in what is called {\it TDU}, carrying to the surface parts of the He-burning ashes (in particular carbon). The lower masses of the interested mass range, i.e. those from about 1.5 M$_{\odot}$ and to about 3 M$_{\odot}$, mix into the envelope enough $^{12}$C to make the C/O ratio become larger than unity (these are the so-called {\it carbon stars}, see a review by Straniero et al. in \cite{str23}).

The temperature achieved at the $ICZ$ base ($T$ ranging, as said, between 2.8 and 3.5 10$^8$ K) is in general sufficient to drive various processes of $\alpha$ capture. In particular, from the $^{14}$N left by CNO cycling after H-shell burning, the chain:

\begin{equation}
^{14}N(\alpha,\gamma)^{18}F(\beta^+\nu)^{18}O(\alpha,\gamma)^{22}Ne
\end{equation} 
can be driven. If the temperature is high enough ($T \gtrsim 3.0 - 3.2 \times 10^8 K$) this is followed by the reaction:
\begin{equation}
^{22}Ne(\alpha,\gamma)^{26}Mg
\end{equation} 
and by the competing:
\begin{equation}
^{22}Ne(\alpha,n)^{25}Mg \label{eq17}
\end{equation} 
with a higher probability for the second reaction. \cred{More specifically, at about 300 MK, the reaction rate ratio $<\sigma v>_{^{22}Ne(\alpha,n)}/<\sigma v>_{^{22}Ne(\alpha,\gamma)}$ is $\sim 3$}. 

Reaction (\ref{eq17}), in particular, makes intense neutron fluxes available, so that for almost two decades it became natural to assume that the s-process was originated in this way, thus accounting for the mentioned observations of Tc by Merrill \cite{mer} in an $AGB$ star. 

The above conclusion seemed to receive a nice confirmation when Ulrich \cite{ul73} noted how the exponential distributions of exposures $\rho(\tau)$, suggested by \cite{seeg65}, might be well accounted for during He-shell instabilities. This can be easily illustrated by considering the neutron exposition $\Delta \tau$ produced by the \neanb reaction in each  $ICZ$ of mass $M_{ICZ}$, overlapping with the subsequent one by a factor $r$ = $\Delta M_{ICZ}/M_{ICZ}$. Indeed, by neglecting the local effect of dredge-up and indicating with $\tau_0$ the ratio $\Delta \tau/(- \ln(r))$, it is simple to derive that:
\begin{equation}
\rho(\tau) \propto \frac{(1-r)}{\Delta \tau} r^{\tau/\Delta \tau} \propto \frac{1}{\tau_0}exp(-\tau/\tau_0) 
\end{equation}
which has the same form as in equation (\ref{eq6}).

The general properties of TP-AGB stars, crossing the phases where {\it thin shell instabilities} are met, were studied by Paczy\'{n}ski in the early seventies \cite{pac70}, \cite{pac71}, and \cite{pac75}, deriving analytical relations linking the main stellar parameters (like the inter-pulse period, the luminosity, the rate of advancement of the H-burning shell), usually expressing them as a function of the H-exhausted core mass, $M_H$. Similar relations are usually easily derived in stellar models, thanks to the convergence of evolutionary tracks to a narrow range of parameter values, ultimately due to the common property of low- and intermediate-mass stars of witnessing only the evolution of a rather thin layer above  a degenerate pre-white dwarf core of a mass below the Chandrasekhar limit.

\begin{figure}[t!!]
\includegraphics[width=\linewidth]{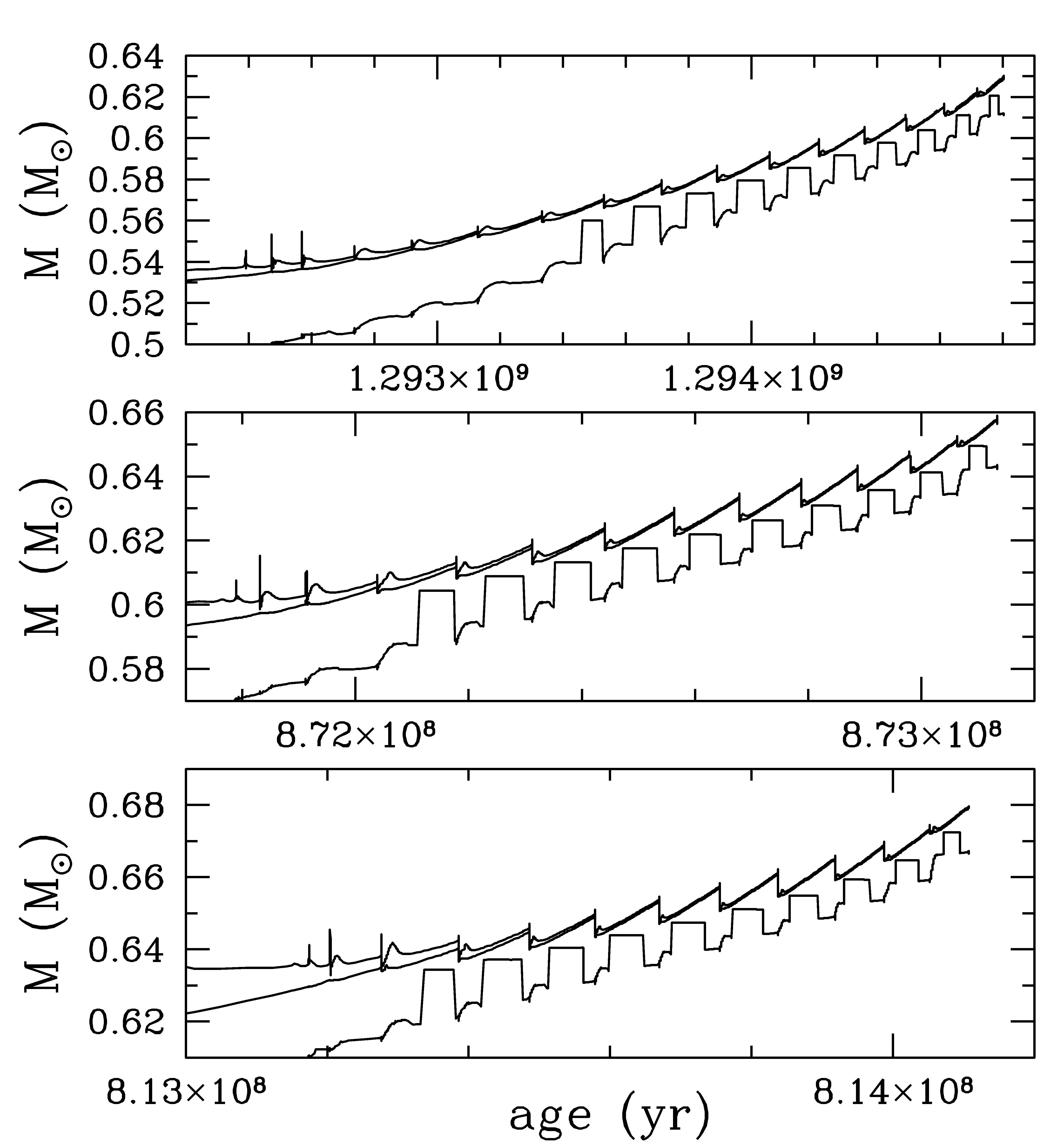}
\caption{The evolution of the positions in mass of the bottom of the convective envelope (M$_{\rm CEB}$), of the H-burning shell (M$_{\rm H}$) and of the He burning shell (M$_{\rm He}$), in models of 2 M$_\odot$, for 3 different metallicity, specifically: solar, i.e [Fe/H] = 0, (top panel), 1/20th solar, i.e. [Fe/H] = -0.03 (middle panel) and 1/200 solar, i.e. [Fe/H] = -0.003 (bottom panel). Note the inward penetration of the convective envelope (TDU) after each thermal pulse and the forward displacement of the He-burning shell marking the activation of the  \ctanb in the inter-shell zone during the inter-pulse period (more details in \cite{cris11}).
}
\label{fig5}
\end{figure}

\cred{The typical structure of the layers outside the CO core, for a low-mass $AGB$ star crossing these evolutionary stages to become a carbon stars (C star) \cite{str23}, is shown in Fig. \ref{fig5}. The figure was obtained with the FuNS code (Full Network Stellar evolution, see \cite{straniero_2006NuPhA},\cite{cristallo_2009ApJ},\cite{cris11})}. Parenthetically, we recall that the notation [Fe/H] refers to:

\begin{equation}
    {\rm [Fe/H]} = \log \left[{\frac{N(\rm Fe)}{N(\rm H)}}\right]_{star}-\log\left[ {\frac{N(\rm Fe)}{N(\rm H)}}\right]_{\odot}
\end{equation}

The fact that s-process-rich C stars observationally tended to appear as being of low mass in contrast with the needs of the \neanb source  was noticed by Icko Iben as early as in 1981 \cite{i81b}, after a decade dedicated by his group to study more massive $AGB$ stars \cite{ti77}, \cite{it78}, \cite{bi80}, \cite{i81a}. As a consequence, he and A. Renzini also suggested a possible way to initiate neutron captures from the alternative \ctanb source
\cite{ir82}, thus pioneering subsequent trends of research.

Initially, the same accounting for TDU episodes, allowing freshly synthesized nuclei to be mixed into the envelope was a serious problem of stellar models, at least for low mass stars ($M \lesssim 3$ M$_\odot$). This was strictly connected to the choice of opacity tables. Negative results on TDU were common, see e.g. \cite{bs881}, \cite{bs882}, while positive outcomes were instead found by  \cite{i75}, \cite{w81} and later by \cite{john87}, \cite{john89}. Here numerical problems (like model inadequacies
e.g. in the mesh rezoning), insufficient opacity tables, and the specific choices for the mixing algorithms at the envelope border all played crucial roles. 

The lengthy calculations required by TDU with the computing facilities then available made the use of semi-analytical models for TP-AGB phases rather common, a method that proved in general effective and provided in the years significant results
\cite{paola96}, \cite{wg98}, \cite{paola22a}, \cite{paola22b}.   

A growing evidence for the mixing of s-process elements to the surface was  in the mean time emerging, after the original Merrill's discovery. Data accumulated rapidly for all the classes of polluted TP-AGB stars, i.e. stars of spectral types $M, MS, S, SC, C(N)$, clarifying the  enrichment in heavy nuclei due to dredge-up along the TP-AGB evolutionary stage (see section \ref{abundances}, and works by e.g. \cite{gus89}, \cite{str23} for reviews widely separated in time).

Before the end of the 1980s, it had become unquestionable that AGB stars enriched in s-elements were of rather low mass, and hence could not have derived their enrichment from neutron captures induced primarily by the \neanb source, due to the too low temperature at the base of TPs in those objects \cite{bus88}. The observational studies that reveal this will be, as anticipated, the object of section \ref{agb data}.

\section{The \texorpdfstring{$^{13}$C}{13C} pocket}
\label{c13_pocket}

As recalled in the previous sections, the TDU is the process responsible for the carbon and heavy-element overabundance commonly observed in the atmosphere of evolved AGB stars. Moreover, TDU itself is the basis for a series of events that culminate in the activation of a powerful neutron source, i.e. the reaction mentioned \ctanb. Indeed, when the H-rich convective envelope penetrates down, the inter-shell zone has already been enriched in \iso{12}{C}, which is the main product of $3\alpha$ combustion during a thermal pulse. Since \cite{str95}, \cite{straniero_1997ApJ}, \cite{gallino_1998}, it became clear that this occurrence is of paramount importance for the synthesis of heavy elements in low-mass AGB stars. 
Practically, during a TDU episode, the H-rich zone from the envelope gets in contact with the carbon enriched inter-shell region, while the H-burning shell is inactive. At that time, if a small amount of hydrogen can diffuse inward, a \iso{13}{C} pocket may form, trough the $^{12}$C(p,$\gamma$)$^{13}$N($\beta^+ \nu$)$^{13}$C reactions. Then, through the activation of \ctan, it provides a very efficient neutron source for the s-process nucleosynthesis. 
Similarly to a sea wave that crashes onto the beach and then recedes, leaving a wet sand zone, the convective envelope, after sinking into the H-poor zone, leaves a layer with a small amount of protons that reacts with carbon to form a \iso{13}{C} reservoir (often called $^{13}C$ pocket. The physical processes shaping the H profile left by the TDU are the matter of a longstanding debate, partly still open, as we outline in the next subsections.

\subsection{The role of convection}
A common feature of all AGB stars is the presence of a deep convective envelope. Usually, the internal convective boundary is limited by the presence of an active H-burning shell, which  implies a local entropy barrier preventing further penetration 
of the convective instability. 
However, this condition is removed after the occurrence of a $TP$. In fact, to compensate for the intense flow of thermal energy generated from the He-flash, the outermost layers expand and cool, causing a shutdown of the H-burning shell. Then, as the convective envelope penetrates inward, reaching the hydrogen-exhausted region, a marked compositional gradient is established at the convective boundary. 
Specifically, the mass fraction of hydrogen  falls abruptly from about 70\% in the convective region to zero  in the underlying radiative zone, where hydrogen was previously burnt.
 This discontinuity in composition induces a sharp variation of the radiative opacity and, in turn, an abrupt change of the temperature gradient. In this condition, the precise location of the regions fully or partially mixed by convection becomes highly uncertain.
Indeed, even a small perturbation that causes further mixing gets amplified
over a dynamical time scale, and the consequent enhancement of the H abundance causes the convective instability to propagate inward. 

The above fact has two main consequences. 
Firstly, it enhances the efficiency of the TDU, allowing a larger amount of carbon and other products of internal nucleosynthesis to be transported into the external envelope.
Furthermore, this event inhibits the growth of the H-ex\-haus\-ted core, hence impacting the star's main macroscopic properties, including its luminosity and the intensity of subsequent $TP$s \cite{straniero_2003pasa}. The observable consequen\-ces are many
and important. For example, the luminosity function of C stars is shifted toward lower luminosities \cite{abia_2020A&A}.

Clearly, a realistic modeling of the convective boundary instabilities during the TDU phase is of paramount importance to obtain accurate AGB star models. This, in turn, affects our understanding of their role in nucleosynthesis.

\begin{figure}
\begin{center}
\includegraphics[scale=0.45]{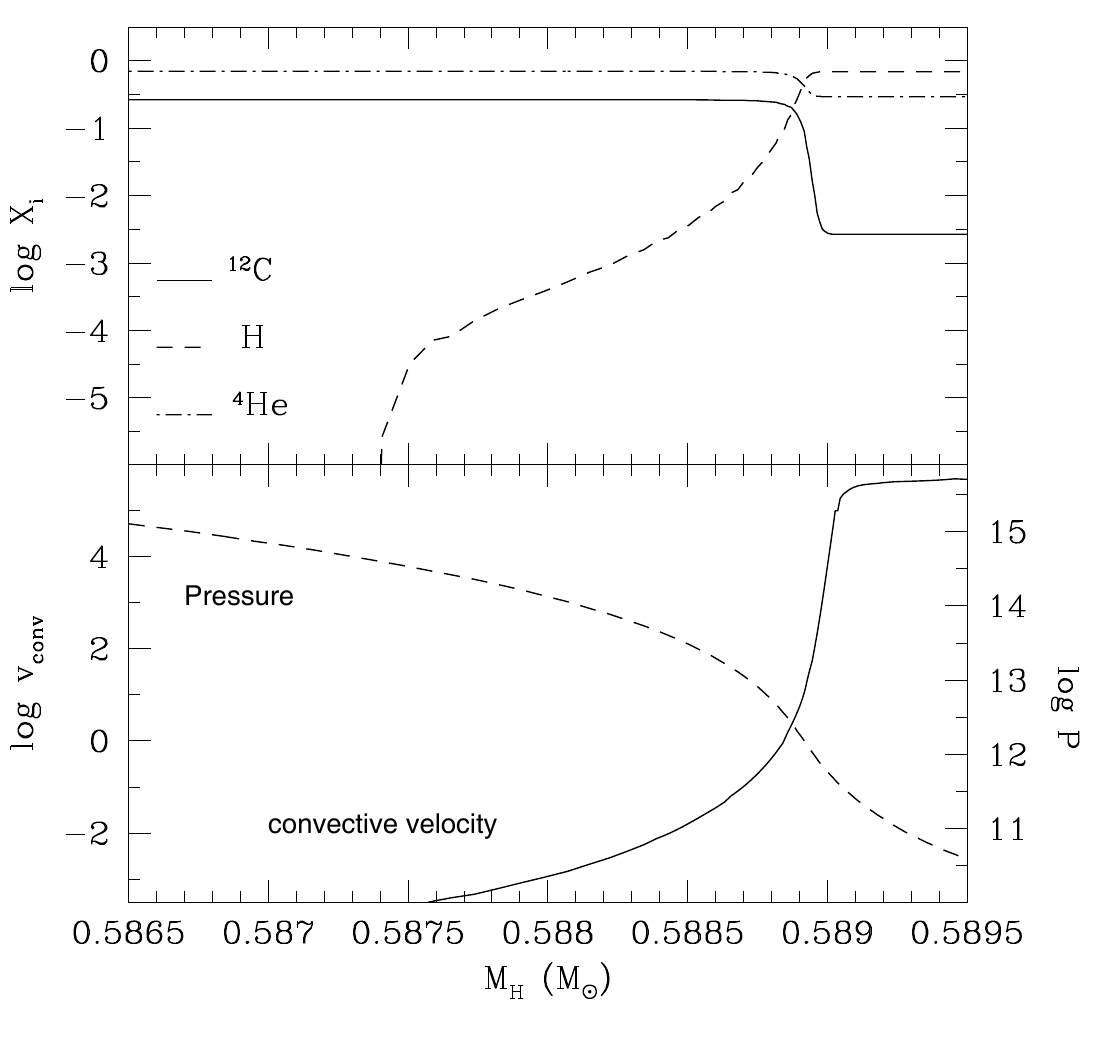}
\caption{The region around the boundary of the convective envelope during the 6th TDU episode of the 2 M$_\odot$ with solar composition (the same model shown in the upper panel of Fig, \ref{fig5}). Upper panel: chemical composition in the transition region between the convective envelope 
and the radiative He-rich zone.
Lower panel: the exponential decline of the convective velocity and the pressure gradient. Here the $\beta$ parameter in equation \ref{vexpo} was set to 0.1. Adapted from \cite{straniero_2006NuPhA}.
}\label{velco}
\end{center}
\end{figure}

Obtaining a final solution to this {\it boundary-mixing problem} would require sophisticated and reliable hydrodynamic tools. So far, only
limited, although instructive, hydrodynamical investigations
have been carried out for such purposes (\cite{freytag_1996A&A}, \cite{herwig_2006ApJ}).
Typically, models confirm that standard mixing-length approaches roughly provide a reasonable description of the physical  properties within a convective zone. They include, in particular, an evaluation of the temperature gradient and of the mean convective velocity. In addition, they also show how convective overshooting, that is, mixing beyond the boundary of a convective unstable zone, can be roughly described by assuming an exponentially decreasing  velocity profile. Specifically:

\begin{equation}\label{vexpo}
    v(r)=v_{cb}\exp(-\frac{|r_{cb}-r|}{\beta H_P})
\end{equation}

\noindent where the suffix {\it cb} indicates quantities calculated at the convective boundary as defined by the Schwarzchild criterion, $H_P$ is the height of the pressure scale, and $\beta$ is a tunable parameter, whose value depends on the specific convective boundary under consideration.
 When this recipe is applied to the convective envelope of an AGB star during the TDU phase (see e.g. \cite{herwig_2000A&A}, \cite{straniero_2006NuPhA}, \cite{cristallo_2009ApJ}, \cite{cris11}), instead of getting a sharp separation between radiative and the convective zones, as is expected for the bare Schwarz\-schild’s criterion, the fully-radiative core and the fully-convective (and homogeneous) envelope turn out to be separated by a rather extended transition zone, where only partial mixing takes place, so that a smooth and stable H-profile may form (see Fig. \ref{velco}).  Within this transition zone, the convective velocity smoothly drops from about $10^5$ cm/s, at the convective boundary, to zero. It was soon recognised that, within a few hundred years since TDU, as H burning restarts, a consistent amount of $^{13}$C may form in the transition layer, through the activation of the $^{12}$C$(p,\gamma)^{13}$N reaction, followed by $^{13}$N decay. 
 
\begin{figure}
\begin{center}
\includegraphics[scale=0.38]{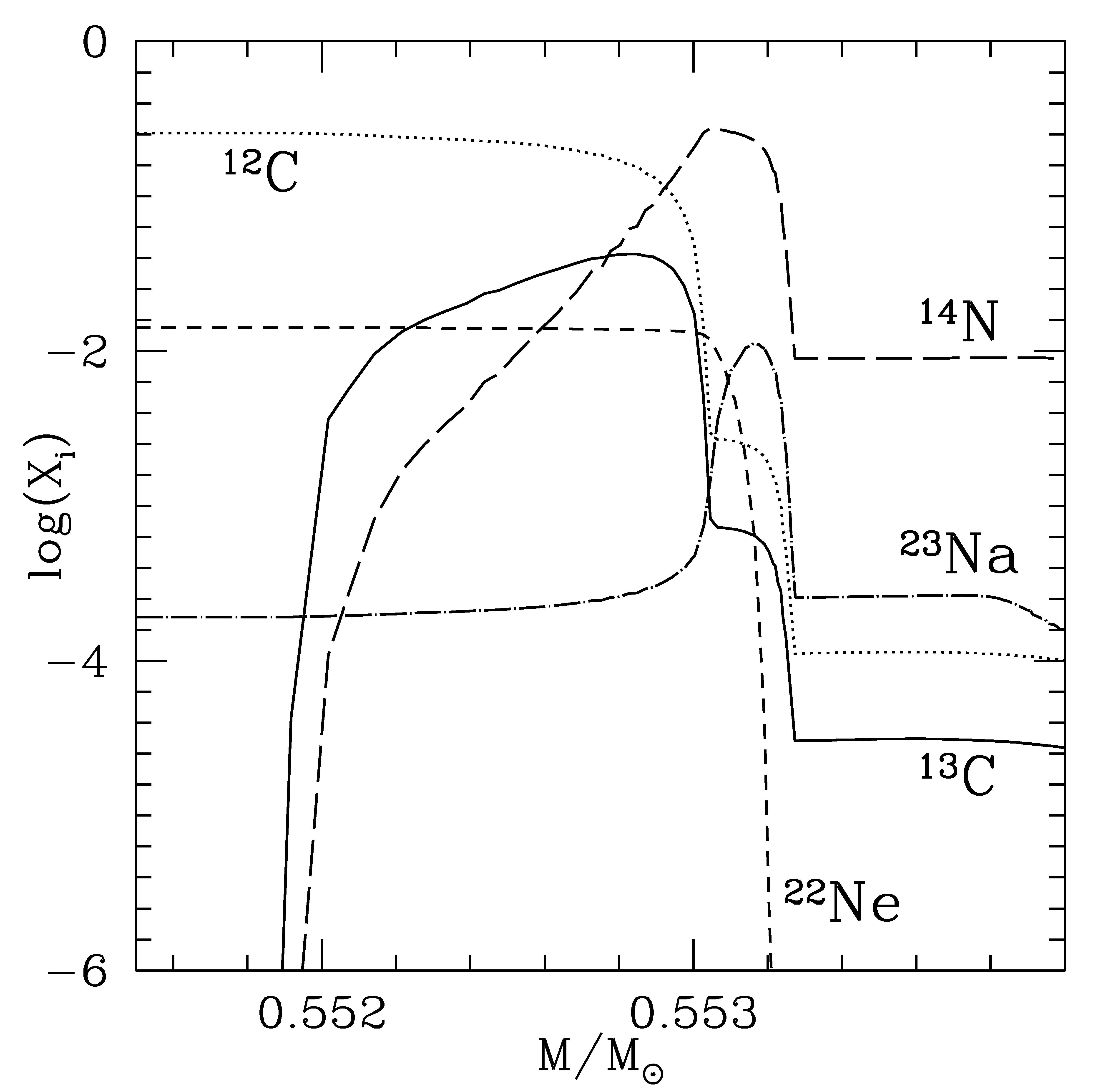}
\caption{The \iso{13}{C} pocket that forms during the inter-pulse that follows the first fully developed TDU episode of the 2 M$_\odot$ model in Fig. \ref{fig5} (upper panel). The mass fraction of the main chemical species are shown. A  well developed \iso{13}{C} pocket starts at $m\sim0.552$ M$_\odot$ and it extends outward for about $10^{-3}$ M$_\odot$. Where more protons were diffused at the time of the TDU (see Fig, \ref{velco}), a \iso{14}{N} pocket, partially overlapping the \iso{13}{C} pocket, forms. Also, note the presence of a smaller \iso{23}{Na} pocket. 
}\label{tasca}
\end{center}
\end{figure}

 A typical \iso{13}{C} pocket thus formed is shown in Fig. \ref{tasca}. It extends for about $10^{-3}$ M$_\odot$ and contains a few $10^{-5}$ M$_\odot$ of \iso{13}{C}, corresponding to more than $10^{51}$ nuclei of \iso{13}{C}. Later on, when the temperature within the pocket becomes $\sim 80$ MK, the \ctanb reaction starts, providing a sufficiently high neutron exposure to synthesize the bulk of the s-process main component. In low-metallicity AGB stars, this also provides the strong component, originally anticipated by \cite{seeg65}. For details on these phenomena and for the dependence of the s-process efficiency on metallicity, see \cite{gallino_1998}.  
 
Provided that the $\beta$ parameter is properly calibrated, this approach allows a reasonable reproduction of the s-process overabundance observed in AGB and post-AGB stars. Also, the yields predicted by AGB star models incorporating the convection treatment described so far provide an excellent reproduction of the solar system abundances of isotopes with $A>90$ that are mainly or exclusively produced by the s-process \cite{prantzos_2020MNRAS}. 
Nevertheless, a deeper analysis of the composition of pre-solar grains, which are thought to originate in the circumstellar envelopes of low-mass AGB stars, suggests that convective processes alone cannot explain their complex isotopic pattern. Hence, some non-canonical processes, usually neglected in extant stellar model calculations, should be invoked to get a better agreement between observations and theoretical predictions \cite{liu_2022Univ}. Let us therefore review in the net subsections the main mechanisms so far explored for such purposes.

\subsection{Rotation}
\label{rot}
Rotation, as well as transport and loss of angular momentum (here after AM), may substantially affect the evolution of low- and intermediate mass stars. Pre-main-sequence stars, such as T Tauri stars, rotate much faster than the Sun, with periods that can be of just a few days.
However, magnetic interactions between the central star and the accretion disc can actually slow down the rotational speed (magnetic braking) or keep it at a nearly constant value. On the other hand, a spin-up may occur during the contraction phase that follows the disc dissipation.

Observations of young stars, those close to the Zero Age Main Sequence (ZAMS), show that the majority  (70\%) of them have rotation velocity $<30$ km s$^{-1}$ with the projected velocity distribution ($v\sin i$) peaked at $\sim8$ km s$^{-1}$ \cite{weise_2010A&A}. Likely, part of the initial AM is lost during the pre-MS evolution. In general, mass loss also implies an AM loss. Moreover, stars with a convective envelope may develop an external magnetic field (and dynamo mechanisms, as in the Sun). Then, the stellar wind is captured by the magnetic field lines out to the Alfv\'en radius. We recall here that the Alfv\'en radius  marks the region where  the magnetic field of an astrophysical object becomes strong enough to affect the flow of plasma in its surroundings. Hence, beyond the Alfv\'en radius, the magnetic field dominates the plasma flow and can prevent material from flowing back freely into the object. The Alfv\'en radius lays well beyond the stellar photosphere. Since the amount of AM lost is proportional to $r^2$, magnetic braking becomes much more significant than it would be achieved by mass loss alone. The Sun offers an excellent example of magnetic braking in an old star with convective envelope and an associated magnetic field. The present equatorial velocity is just 2 km s$^{-1}$, well below the average of younger main-sequence stars, but from the analysis of acoustic oscillations it results that the radiative solar interior rotates rigidly and faster than the convective envelope (see e.g. \cite{howe_2009LRSP} for a review of solar rotation). Hence, an efficient transport of AM and the consequent redistribution of the angular velocity between the core and the envelope, can increase the total AM loss. 

A primary AM transport mechanism in stars is convection itself. In practice, mixing within convective regions results in nearly uniform rotation (solid-body rotation). However, even in the radiative zones of a star, rotation-induced instabilities and magnetic fields may be responsible of a forward AM transport. In a rotating star the isothermal surfaces do not coincide with the equipotential ones, due to the ellipsoidal geometry induced by rotation itself. This fact is well known to result in a thermal instability (called {\it Eddington–Sweet, or ES, instability}), causing a slow circulation of matter along meridians, from the equator to the poles and vice versa (meridional circulation). Moreover, in case of a differential rotation, other instabilities may occur.  Among them, noticeable is the Goldreich–Schubert–Fricke (or GSF) instability, which arises in the case of a steep specific AM gradient. Eventually, magnetic fields provide a torque both on stellar surface (via stellar wind) and internally (via the so-called Maxwell stresses), depending on interactions between electromagnetic forces and mechanical momentum. All this acts as a form of super-viscosity, tending to uniformize the rotational profile. 

Incorporating rotation effects into a stellar evolution code is not straightforward. The reason is that most of these codes assume spherical symmetry (1-D). Nonetheless, in case of moderate rotation velocities it is possible to treat the resulting deviations from the spherical symmetry by appropriately modifying the energy transport equation and adding the centrifugal force to the hydrostatic equilibrium equation \cite{endal_1976ApJ}. Hence, like with convective mixing, the AM redistribution and the associated mixing can be described as diffusive processes. This phenomenological procedure requires appropriate criteria to determine the zone where the various rotationally-induced instabilities are active and  the calibration of a number of free parameters entering the calculations of the diffusion coefficients. 

The effects of rotation on s-process nucleosynthesis in low-mass AGB stars have been investigated by various authors \cite{herwig_2003ApJ},\cite{siess_2004A&A},\cite{piersanti_2013ApJ},\cite{hartogh_2019A&A}.
These studies demonstrated that rotation alone should not influence the formation of the \iso{13}{C} pocket at the time of TDU. The duration of TDU is actually too short to allow rotationally-induced mixing to modify the H profile remaining after the convective envelope has receded  (see previous section). 
On the other hand, during the longer inter-pulse phase, once the \iso{13}{C} pocket is set on, ES and GSF instabilities are active in the inter-shell region (see Fig. 7 in \cite{piersanti_2013ApJ}). The effects of the resulting mixing is twofold. 

\begin{figure}
\includegraphics[width=\linewidth]{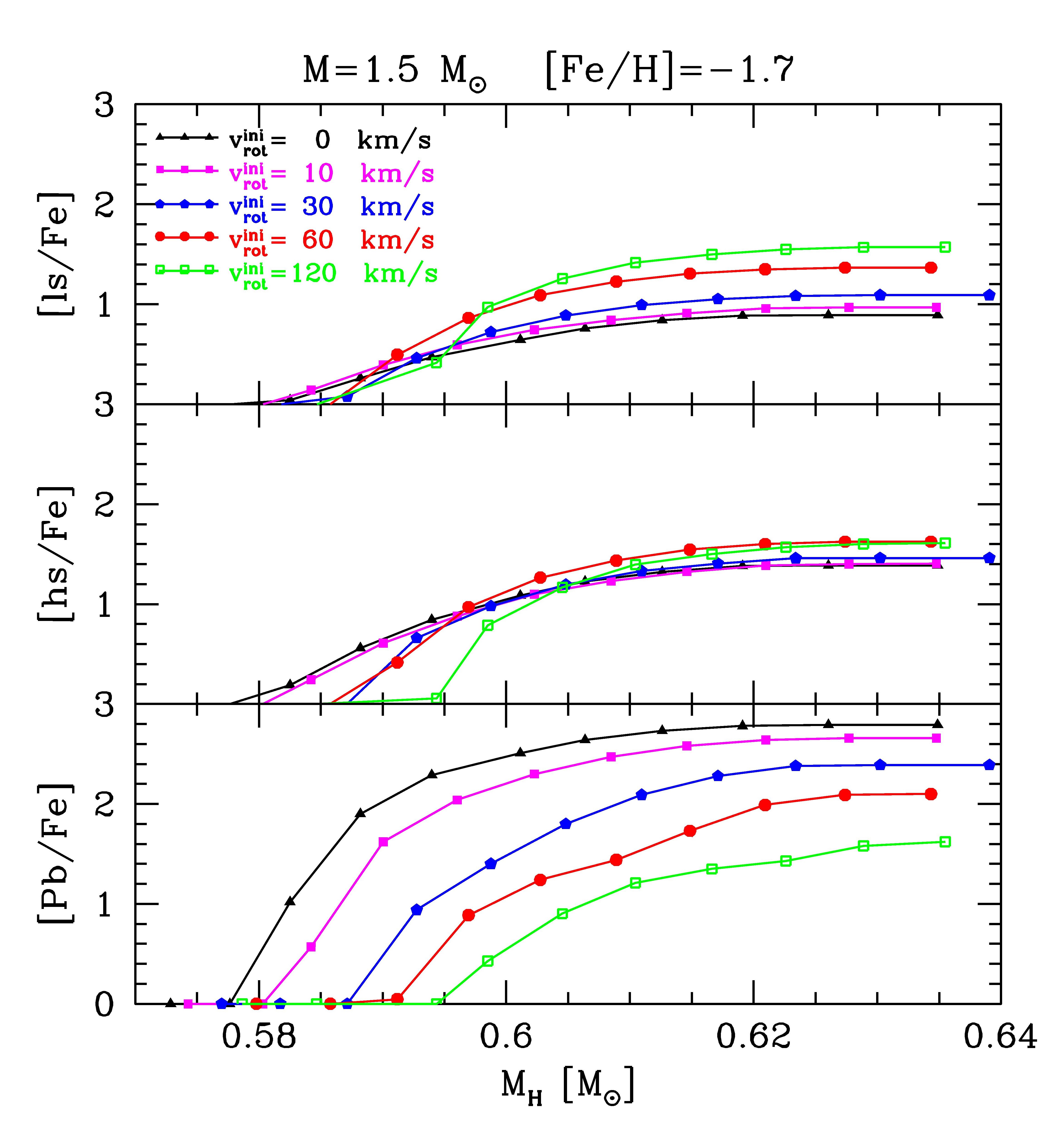}
\caption{Evolution of the parameters [ls/Fe], [hs/Fe] and [Pb/Fe] (see definitions in the text) as a function of the core mass for of a 1.5 M$_\odot$ model with [Fe/H]$= - 1.7$. Each curve refers to a different initial rotation velocity. Adapted from \cite{piersanti_2013ApJ}. 
}\label{rota2}
\end{figure}

Firstly, the  \iso{13}{C} pocket is stretched, that is the extension of the pocket increases while the average \iso{13}{C} mass fraction is lowered. At the same time, some \iso{14}{N} is moved from above into the \iso{13}{C} pocket, so that the effective \iso{13}{C} mass (i.e. m$_{\rm{^{13}C}}-$m$_{\rm{^{14}N}}$) is reduced. 
Indeed, \iso{14}{N} is a strong neutron poison, so that an increase of its abundance in the \iso{13}{C} pocket should leave less neutrons for the synthesis of elements beyond iron. A comparison of the resulting heavy element abundances as obtained by varying the initial rotation velocity is shown in Fig. \ref{rota2}. The Fig. adopts a very common representation for indicating average abundances around the main s-process peaks. [ls/Fe] represents such an average as compared to Fe (scaled to solar values as in equation (19)) for elements near Sr (including in this case Sr, Y and Zr; but one must be careful that the admixture of elements is not always the same in all publications). Similarly, [hs/Fe] indicates the analogous average abundance for elements at the second peak, i.e. near Ba. In this case the chosen elements are Ba, La, Ce and Nd. As a comparison, the same logarithmic solar-scaled abundance is shown for Pb, at the third peak.

The most relevant effect of rotation is the reduction of the abundances of the heaviest isotopes produced by the s-process. In particular, the large  contribution  from low-metallicity AGB stars to the solar Pb is substantially suppressed for fast-rotating models. 

In any case, only marginal effects of rotation on s-process nucleosynthesis are expected for the majority of low-mass stars AGB stars, for which the initial rotation velocity is $<30$ km s$^{-1}$. Moreover, it should be recalled that current rotating AGB stellar models do not include the mentioned angular momentum loss associated to magnetic braking.
As stressed by \cite{hartogh_2019A&A}, asteroseismic observations of red clump stars show that their cores rotate more slowly than predicted by current models, likely due to the neglected effects of magnetic fields. This means that the impact of rotation alone on nucleosynthesis probably remains marginal, even at higher initial velocities (possibly as large as 100 km s$^{-1}$). On the other hand, as discussed in the next section, a possible  coupling of rotation and gravity waves may enhance turbulence below the bottom of the convective envelope at TDU time, thus inducing a more efficient diffusion of protons in the underlying radiative region \cite{talon_2007ASPC}.  
 
\subsection{Gravity waves}
\label{gravity}
Internal gravity waves (IGW) are generated  when a fluid element is displaced vertically and the surrounding fluid, more or less dense, pulls it back, causing it to oscillate. This is the same principle behind waves on the ocean's surface. Gravity waves are also commonly observed in the atmosphere of our planet.
In stars, IGWs are usually generated near the border of a convective region. The turbulent motions of convection excite the waves, which then propagate into the stably stratified radiative layers. In Sun-like stars, IGWs, which are supposed to be generated at the base of the convective envelope, propagate inward, down to the core.  In more massive stars, IGWs can also be excited at the boundary of the convective core and propagate outward.

IGWs may transport angular momentum between the fast-spinning core and the slower-rotating outer layers. As explained in the previous section, magnetic braking slows the surface rotation of the Sun, so that the core should spin much faster than the outer layers. However,  \cite{talon_1998AA}  found that IGWs generated at the bottom of the convective envelope of the Sun, coupled to meridional circulation, can efficiently redistribute angular momentum, leading to a uniform rotation profile. Also, high-resolution hydro-dynamical simulation confirm the efficiency of angular momentum transport by IGWs (see, e.g. \cite{rogers_2013ApJ}). This occurrence reconciles the theoretical expectations with the helioseismology measurements of solar internal rotation.

IGWs can also account for non-convective mixing in stellar interiors. This phenomenon could account for the nitrogen enhancement observed in massive main-sequence stars \cite{brinkman_2025A&A}. On the other hand, mixing induced by IGWs has been also invoked to explain anomalous overabundances of lithium observed in some evolved red giant branch, red clump and AGB stars.
The mechanism responsible for this (extra-mixing)  may be the dissipation of IGWs. When the wave generated by turbulence in the convective envelope of an AGB star propagate into the underlying stable region, it passes through layers with increasing density, so that its amplitude changes and may become unstable. According to \cite{denissenkov_2003MNRAS}, this occurrence generates turbulence and, in turn, mixing of the surrounding fluid. Instead, \cite{talon_2007ASPC} (see also \cite{talon_2006EAS}) proposed an even more efficient mechanism, for which waves act on the rotation profile that becomes locally unstable and produces mixing.  In both cases, this extra-mixing can move downward protons from the convective envelopes, providing a promising mechanism for the generation of the $^{13}$C pocket during the occurrence of a TDU episode \cite{battino_2016ApJ} (see also \cite{battino_2019}).

\subsection{Magnetic circulation}
Although we have overwhelming evidence, since several decades, of stellar magnetic phenomena occurring both in the Sun (see e.g. \cite{mest}) and in various classes of active stars (see e.g. \cite{rod82}),  supported by measurements over a wide range of wavelengths from the X-ray to the radio domains, it was not straightforward to assume that they could be present also in AGB stars. Actually, the  atmospheres of such extremely cool red giants rotate so slowly that X-ray observations (e.g. from ROSAT, XMM and CHANDRA) in general excluded the formation of coronae in them (see e.g. \cite{ayres},\cite{hun}).

For many years through the nineties of the XXth century, the clear existence of a {\it coronal 
dividing line} in the HR diagram, to the red side of which (and including the upper part of the
Giant Branch) no X-ray emission was observed from stellar atmospheres, induced researchers to assume that magnetic fields had only marginal, if any, effects in TP-AGB stars. 

\begin{figure}
\includegraphics[width=\linewidth]{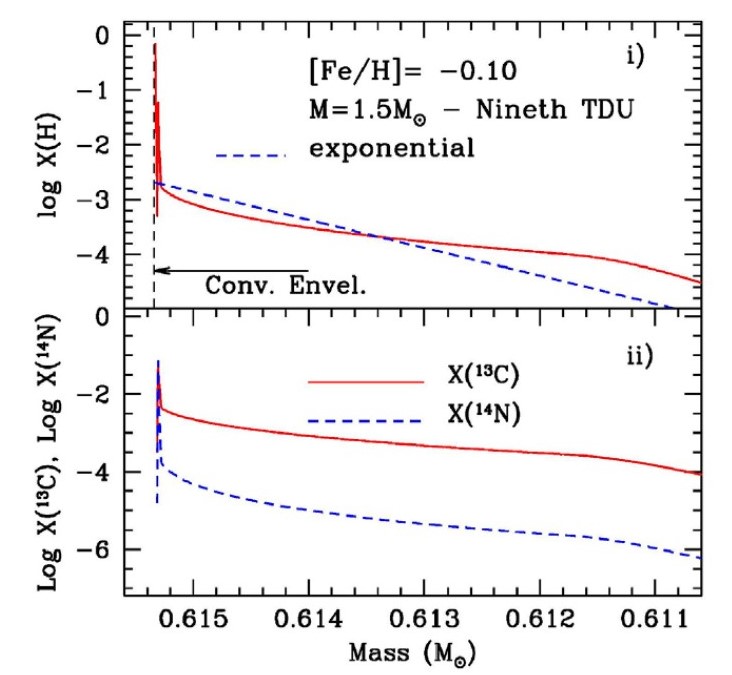}
\caption{An example of the proton pocket (upper panel) that results from magnetic buoyancy at TDU, inducing the penetration of material from the envelope into the He-rich layers. When subsequently the H shell is reactivated, rapid occurrence of the $^{12}$C(p,$\gamma$)$^{13}$N($\beta^+ \nu$)$^{13}$C(p,$\gamma$)$^{14}$N mainly makes fresh $^{13}$C available for $\alpha$ captures (generating $^{16}$O and neutrons, before the development of the next ICZ). The small proton concentration permits only very little production of $^{14}$N (bottom panel).  Then, in the ICZ a small amount of $^{22}$Ne, formed from $^{14}$N($\alpha$,$\gamma$)$^{18}$F($\beta^+\nu$)$^{18}$O($\alpha$,$\gamma$)$^{22}$Ne releases a second (very small) flux of neutrons.} 
\label{magpocket}
\end{figure}

However, more recently it was noticed by \cite{hol} that the lack of X-ray observations  from coronae might simply indicate the burial of magnetic flux tubes inside the huge convective envelopes. This consideration later induced \cite{bus+07} to advance the suggestion that such a burial might promote mixing, and actually induce the formation of a $^{13}$C pocket at the H-He interface during the TP phase. Subsequently \cite{nb}, adopting a very simple geometry, derived 
an exact 2D and 3D solution for the MHD equations, indicating an efficient process of mixing during the double-shell phase of AGB stars, for reasonable values of the magnetic field (i.e. $\sim 10^5$ G, similar to the field found in flux tubes at the base of the solar convective envelope). Their conclusions were rather similar to those advanced, more qualitatively but  many years before, by \cite{park}, as consequences of magnetic buoyancy.
Extensive calculations were then performed to verify the suggestions by \cite{nb}. They included works by \cite{ves+20}, \cite{ves+21}, \cite{b+21}, \cite{mag+21}, \cite{pal+21}, \cite{tai+22}, \cite{ves+22}, \cite{ves+22b}, \cite{b23}, \cite{ves23} and various other attempts. The general results of such models and applications induced the rather firm conclusion that magnetic buoyancy can indeed provide a valid mixing mechanism, yielding the production of $^{13}$C reservoirs below the H-He interface driving neutron capture phenomena adequate to account for several constraints posed by AGB and post-AGB  stars, as well as by the Galactic enrichment of elements produced by the s-process and by measurements of trace isotopes in presolar grains. 
A rather exhaustive comparison between s-process and r-process predictions for nuclei in the range from Sr to Ce was performed by \cite{b+22}. They provide the most significant agreement so far available between expectations for the s- and r-components of the solar abundances.  A typical example of a $^{13}$C pocket formed by 
magnetic mechanisms and used in such comparisons is shown in Fig. \ref{magpocket}. From the figure one can immediately see  the most peculiar feature of the mixing mechanisms adopted. These last are suitable to limit the production of $^{14}$N to a very low level. 

The need for such a property was early inferred, through parametric computations of s-processing, by comparisons with the correlations of Sr and Ba isotopes in presolar SiC grains from the Murchison meteorite by \cite{liu}. 

There are three specific (obvious) properties of this type of $^{13}$C reservoir with low $^{14}$N. They descend from the fact that $\alpha$, $p$ and $n$ captures on $^{14}$N itself are sharply reduced. This will, as a first thing, reduce remarkably the efficiency of the reaction channel that from $^{14}$N leads to fluorine production. Essentially, nitrogen will consist only in the rather small abundance that is, in any case, produced by the recycling of $^{12}$C (which is produced locally, mixed by TDU into the envelope and then included into the intershell zones by the proceeding of the H shell). The consequences of this reduction of the available $^{14}$N abundance on the nucleosynthesis of $^{19}$F were verified by \cite{ves+22} and found in agreement with the requirements of the galactic enrichment of this element. Another general property induced by the limited presence of $^{14}$N 
is the virtual absence of competition to s-processing by the so called {\it light neutron poisons}, mainly coming from nuclesoynthesis channels that produce light nuclei again starting from $^{14}$N. The third and probably most important effect is a very small production of $^{22}$Ne from the chain: $^{14}$N($\alpha$,$\gamma$)$^{18}$F($\beta^+\nu$)$^{18}$O($\alpha$,$\gamma$)$^{22}$Ne, so that the nucleosynthesis 
expected in the ICZ phase is sharply reduced.

\section{Crucial observational data}
\label{abundances}
\subsection{Normal AGB star observations}
\label{agb data}
 The study of abundances in AGB star atmospheres gives insight into their internal structure and
 mixing processes, hence it is of paramount importance. However, the spectrographic work on them is one of the most difficult
 tasks in the field on stellar abundances. The main reason for this difficulty is the crowdedness of the spectra,
 which makes the synthesis of the relevant spectral regions from model atmospheres a mandatory task. 
The spectrum calculation  requires a large set of atomic and molecular data because thousands of spectral lines may affect the spectral  features of interest. Furthermore,  AGB stars are variable objects and have dynamical atmospheres, yet  most of the abundance work was (until recently, see \cite{gus07}) based on static model atmospheres. This means
 that the abundance results may still suffer from considerable systematic errors. 

In spite of the above difficulties, chemical abundances for s-elements have been derived for AGB stars during the last three decades. Comparisons with theoretical predictions shows a comfortable agreement,  confirming that most of the s-element enhanced AGB stars are of low mass ($\leq 3$ M$_\odot$), and that the main source of neutrons is the $^{13}$C$(\alpha,n)^{16}$O reaction.

Here, we'll give some examples of the above long activity, especially for pioneering works in this field. Obviously, still now the most relevant observational result remains the detection of the unstable s-element Tc by Merrill \cite{mer} in the spectrum of some AGB stars, which (as mentioned before) is a compelling evidence for the operation of the s-process in the interior of such evolved Red Giants.

\begin{figure}[t!!]
\includegraphics[width=\linewidth]{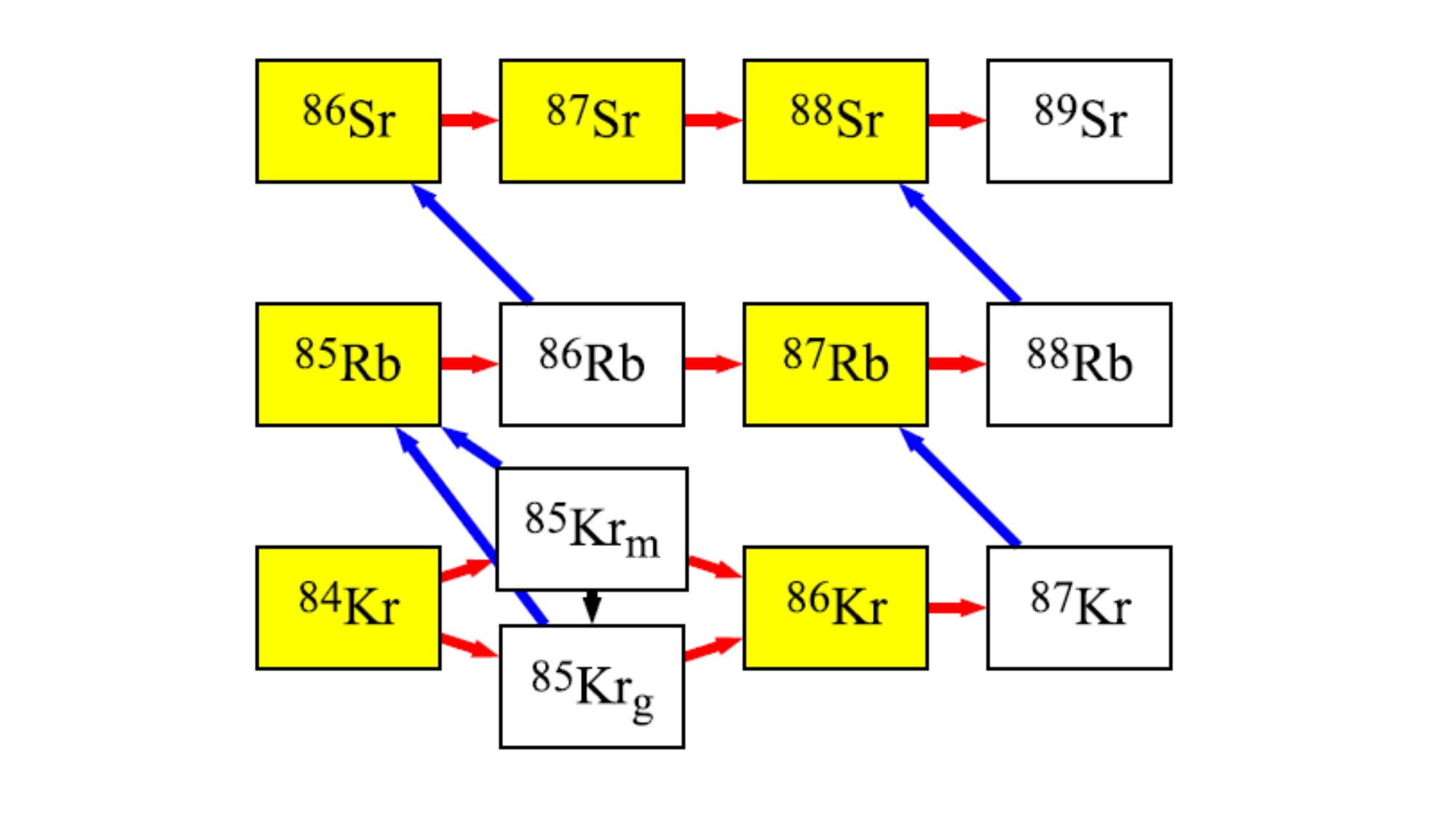}
\caption{The s-process path through the unstable nucleus $^{85}$Kr, with 36 protons and 49 neutrons, i.e. very close to the neutron magic number N = 50. The figure shows the various branches followed by the reaction flow, as discussed in the text.}
\label{fig8}
\end{figure}

Concerning the abundant class of O-rich AGB stars, the Texas group by Smith \& Lambert (\cite{smi85}, \cite{smi86}, \cite{smi90}) determined the abundances of a
number of metals (especially close to the iron peak, e.g. Ti, Fe, Ni) and of s-elements (Sr, Y, Zr, Ba, Nd) in M, MS and S stars from spectra in the near infrared
(which are less crowded than in the visual band). The heavy-element abundance distributions found in such works, together with those observed in the binary CH giants and Ba stars (probably polluted by a now evolved AGB companion) by \cite{van92} and \cite{ror21}, also strongly suggested that the s-process abundance enhancements had been 
created by neutron flows coming from the $^{13}$C$(\alpha,n)^{16}$O neutron source. An important confirmation that these stars have low mass came, in particular, from the analysis of elements near the s-process branching point at $^{85}$Kr. This was also recently considered
by \cite{tann}, who stressed the importance of the $^{85}$Kr$^m$ isomeric state for the synthesis of the long-lived radio-nucleus $^{87}$Rb.
In fact, the peculiarities of the neutron flow around the magic neutron number N = 50, and through the unstable isotope $^{85}$Kr in particular ($N = $49), offer the possibility of deriving an observational estimate for the neutron density. Neutron captures on $^{84}$Kr lead to the 10.7 yr ground state of $^{85}$Kr or to the 4.5 h isomeric state $^{85}$Kr$^m$ at 350 keV. This last $\beta$-decays by roughly
80\% to $^{85}$Rb, while the remaining 20\% undergoes an internal $\gamma$-transition to the ground state of $^{85}$Kr. The production rate of
$^{85}$Kr$^{m}$ is quite important, a\-moun\-ting to about 50\% of the total neutron capture flow, as shown by 
cross sections of $^{84}$Kr. The half-life of the ground state of $^{85}$Kr is long enough for the s-process flow to proceed further
to $^{86}$Kr and then to $^{87}$Rb, in a competition with $\beta$-decays to $^{85}$Rb that depends on the neutron density. Details of the reaction network around $^{85}$Kr are illustrated in Fig. \ref{fig8}. 

When the neutron density becomes higher than a few 10$^8$ cm$^{-3}$, the neutron channel is open and the s-process path feeds the neutron magic nuclei $^{86}$Kr and $^{87}$Rb. Since $^{85}$Rb has a neutron capture cross section larger by a factor of $\sim 10$ than $^{87}$Rb, the total abundance of Rb, as well as its isotopic mixture, are both very sensitive to the neutron density. Indeed, the total Rb abundance can differ by one order of magnitude, depending on whether the low- or the high neutron density routing is at work. Therefore, the relative abundance of Rb to other elements in this region of the s-process path, such as Sr, Y, and Zr, can be used to estimate the average neutron density of the s-process and, as a consequence, to infer the mass of the star. Actually in IMS, where the temperature at the base of ICZs can reach and overcome 30 keV, neutrons would maily be made available by the \neanb source at high values of $n_n$. In fact, \cite{lam95} used the ratio [Rb/Sr] in their sample of M, MS and S stars to estimate the neutron density $n_n$ at the time of s-processing. This was again found to be consistent with lower values of this parameter, confirming that the operation of s-processing had to be driven by the reaction \ctanb, which fact was suggestive that the observed AGB stars were of low mass.

Concerning AGB carbon stars (spectral type N), the first detailed chemical analysis of s-elements was performed by Utsumi
\cite{uts70}, \cite{uts85}.  He found that N-type stars in the Galaxy were typically of solar metallicity, presenting mean s-process 
element enhancements by a factor $\sim 10$ with respect to solar. Later and more accurate studies, based on high-resolution and high signal-to-noise spectra in an extended sample of N stars, by Abia et al. (\cite{abi01}, \cite{abi02}), confirmed such stars to be of nearly solar metallicity, but revised the production factors, on average, to $<$[ls/Fe]$>=+0.67\pm 0.10$ and $<$[hs/Fe]$>=+0.52\pm 0.29$ (for the definition of these parameters we refer so sectieon \ref{rot}, in particulr to the caption of figure \ref{rota2}). These estimates were about a factor of two lower than in Utsumi's works. 

The enhancement values thus found are of the same order as those derived in the less evolved S stars. Abia et al., in the papers \cite{abi98}, \cite{abi01}, also repeated the analysis of the $^{85}$Kr-branching in carbon stars and reached essentially the same conclusions as \cite{lam95}, using the [Rb/Y,Zr] ratios that they re-determined for their new observations: AGB carbon stars had to be of low mass.

As mentioned above, a crucial parameter in s-process nucleosynthesis is the neutron exposure. An estimate of the average value of this parameter was determined observationally through the [hs/ls] ratio. As this represents the relative production of the second to the first s-process peak, it is expected to increase with increasing exposure and decreasing stellar metallicity. This is because the production of $^{13}$C  is of $primary$ origin, i.e. depends on H and He originally present in the star (converted into $^{12}$C through the triple$-\alpha$ reaction). On the contrary, the abundance of Fe, the main seed in the s-process, depends on the initial metallicity, i.e. is of a $secondary$ nature. Decreasing [Fe/H] results in almost the same amount of $^{13}$C and neutrons produced, but also in a lower amount of Fe. The overall result is that the number of neutrons captured per each iron seed gets larger and the s-process can easily reach the second peak. Below [Fe/H]$\sim -1.0$, the third peak can also be achieved and the [hs/ls] ratio becomes roughly constant as the first and second peaks reach a steady-state equilibrium.
This analysis of the  [hs/ls] ratio versus metallicity was done by \cite{abi02}, \cite{del06}, \cite{abi08}, comparing the observational results for AGB carbon stars at different metallicities with  theoretical predictions  of s-process  nucleosynthesis obtained by meanss of the FuNS code \cite{cris11}. Globally, theory and observations are in fair agreement, despite a large observational uncertainty. Nevertheless, for each choice of [Fe/H] there is clearly a spread in the data, well beyond the observational errors. The same spread exists in the models for different ranges of the main parameters. This spread may be real and has been ascribed to different amounts of $^{13}$C burnt during the interpulse phases, depending on several factors: mass of the star, metallicity, rotation, diffusive mixing, strength of magnetic fields etc. The same result was recently found in Ba-stars \cite{ror21}, which again supports the accepted framework of the s-process in low mass AGB stars. The s-elements at the second peak are preferentially produced with decreasing metallicity, with respect to the first peak, and $^{13}$C$(\alpha,n)^{16}$O remains the main source of neutrons.   

\subsection{Post-AGB stars} 
\label{post}

Strong enrichment in both light and heavy s-elements (the mentioned $ls$ and $hs$ parameters) is consistently observed in post-AGB objects belonging to the Milky Way and the Magellanic Clouds (MCs) \cite{her05,bu99}. However, while theoretical models predict that $[\text{hs/ls}]$ should increase with decreasing metallicity, observations do not always show a correlation between $[\text{Fe/H}]$ and $[\text{hs/ls}]$ in post-AGB stars (see, e.g.\cite{Des16,Des15}).

\cred{Noteworthy, in post-AGB stars, the ls index is often computed as the mean abundance of Y and Zr, but not Sr, while the hs index commonly averages La, Ce, and Nd, excluding Ba. The reason is that Sr and Ba lines, usually resonance lines or lines with very low excitation energies, are very strong in the spectra of these stars and, in turn, saturated \cite{Des16} \cite{Des15}. The same issue is also encountered for the coolest carbon stars \cite{abi02}.}

The Pb peak marks the termination of the neutro$n$-capture chain process, as $^{208}\text{Pb}$ is a doubly magic nucleus. Nucleosynthesis models generally predict that low-metallicity AGB stars should display large Pb overabundances relative to other s-elements, owing to increased neutron exposure per iron seed nucleus.
However, for the majority post-AGB stars, only upper limits for the Pb abundance have been reported. These values, typically derived from the Pb I 4057.807 Å line through spectral synthesis, often fall significantly below theoretical predictions, particularly for stars with $[\text{Fe/H}] < -0.7$ \cite{Des16}.
Among the few objects with reliable Pb detection is J003643 in the Small Magellanic Cloud, although even its relatively large abundance of $[\text{Pb/Fe}] = 3.16 \pm 0.21$ remains lower than predicted by current AGB models, which reproduce the abundances of the lighter s-elements \cite{Menon25}.

Hence a major challenge in reproducing post-AGB observations lies in the task of s-process models to simultaneously match the observed abundance patterns beyond the Ba peak (e.g. Eu, Er and Hf) while also satisfying the constraints on Pb.
This issue arises not only for post-AGB stars but also for Solar System abundances (see Sect. 3), as the nucleosynthesis pathway from the $N=82$ shell closure (Ba to La) to $^{208}\text{Pb}$ involves numerous branching points caused by unstable nuclei and isomeric states. For many of these last, neutro$n$-capture cross sections and half-lives under stellar plasma conditions remain poorly determined. Consequently, uncertainties arise from both nuclear physics and stellar models, including the treatment of neutron sources and convective or non convective mixing \cite{bus22}.
Over the years, several efforts have been made to allow standard $^{13}\text{C}$ pocket models to reproduce the observational constraints. These range from classic fine-tuning of parametric models (see, e.g. \cite{tri16}) \cred{to more exotic scenarios in which protons ingested during a thermal pulse  drive intense neutron fluxes, with neutron densities much higher than typical of the s-processing. The resulting nucleosynthesis episodes are generally referred to as the $i$-process (from $intermediate$, name assigned to a range of neutron density in  between the traditional s- and r-processes, $n_n \sim 10^{13}$–$10^{15}$ cm$^{-3}$). Whether or not such a i-process can be active in different sites, including low-metallicity AGB stars and post-AGB stars undergoing late thermal pulses (\cite{cristallo_2007ApJ} \cite{cristallo_2009PASA} \cite{straniero_2011ASPC} \cite{hervig_2011ApJ} \cite{lug15} \cite{denissenkov_2017ApJ} \cite{choplin_2021A&A} \cite{wiedeking_2025NatRP}), is a matter of debate, as well as its potential contribution to the galactic chemical evolution.}

\subsection{Pre-solar grains}  
With the term $presolar$ $grains$, we intend tiny dust crystals of C-rich (e.g. SiC) or O-rich (e.g. Al$_2$O$_3$) compounds so resistant to chemical, $\gamma$-ray and mechanical solicitations that they can be formed in stellar winds or in the outer envelopes of evolved stars and survive all the stresses of a travel through the interstellar medium and of the inclusion into the presolar cloud. They are now found as tiny components (parts per million) in certain primordial meteorites (especially carbonaceous chondrites) and probably represent the most precise constraint we have on models of s-process nucleosynthesis, although they may need complex procedures to be recovered (see e.g. \cite{zin}, \cite{zin0}, \cite{zin1}, \cite{zin2}).

The analysis of such grains, mainly through very sophisticated mass spectrometry techniques suitable for nano-particles, allows us to trace back the isotopic composition of the stellar envelopes in which they formed. 
Among presolar grains, the so-called  mainstream (MS) SiC grains are the largest sample available. They show trace abundances of s-process nuclei (such as Sr, Zr, Mo and Ba). For this reason, they are believed to have formed in carbon-rich AGB stars, in which the main source of neutrons for the s-process is the \ctanb reaction.
The isotopic abundance ratios in the grains can be determined with high precision (note that normal stellar spectroscopy provides only atomic abundances, with very little chance of obtaining isotopic admixtures, e.g. through molecular lines). As some of the isotopic abundances reflect the neutron-to-seed ratios in the nucleosynthesis site, the comparison between grain data and stellar models allows us to infer important information about the $^{13}$C pocket and the possible mixing mechanisms responsible for its formation.

In the 1990's the studies of MS-SiC compositions have been deeply coupled with the investigation of s-process nucleosynthesis (see e.g. \cite{gal-gr1}, \cite{gal-gr2}, \cite{gal-gr3}, \cite{gal-gr4}, \cite{gal-gr5}). Today, recent modeling that includes magneto-hydro\-dynamic (MHD) mixing mechanisms for the formation of the $^{13}$C pocket, has successfully reproduced the average isotopic composition of heavy elements in these grains, confirming that low-mass AGB stars (1.5 to about 3 M$_{\odot}$) are probably their dominant progenitors \cite{pal21,ves20}.

The abundances of nuclei near the magic neutron numbers N=50 (Sr to Zr) and N=82 (especially Ba) are extremely sensitive to the neutron exposure established by the reaction $^{13}$C($\alpha$,n)$^{16}$O. Due to the presence of important branching points near the above neutron magic numbers, some of the isotopic ratios can be shifted in favor of one isotope or the other, depending on the neutron flux to which the seed nuclei are exposed \cite{liu-Ba}. 

The analysis of the isotopic composition of solid materials is often presented in terms of per-mil shifts with respect to solar values, defined as $\delta_{i,j}$ for nuclei -i and -j, being:
\begin{equation}
\delta_{i,j} = \left[\frac{(\frac{N_i}{N_j})_{sample}}{(\frac{N_i}{N_j})_{\odot}}-1\right]\times 1000
\end{equation}
Several MS SiC grains from the Murchison primitive meteorite (so named from the Australian village where it fell in 1969) revealed a strong depletion in the $\delta\rm{(^{138}{Ba}/^{136}{Ba}})$  ratios, down to below $-400$\textperthousand. 
As illustrated by \cite{liu-Ba}, these severely negative values are difficult to be accounted for by parametric post-process models, even with an ad-hoc zoning of the $^{13}$C reservoir, such as those presented in \cite{liu}. 
In order to successfully model these observed isotopic ratios, a peculiar $^{13}$C pocket profile is needed. The latter was achieved by the cited authors by adopting a simplified three-zone division of the pocket, with the second layers extremely extended.

However, analysis of the correlated ratios  $\delta\rm{(^{88}{Sr}/^{86}{Sr})}$ and $\delta\rm{(^{138}{Ba}/^{136}{Ba})}$ revealed that also the total mass of the $^{13}$C pocket is involved,  with values in excess of $10^{-3} M_{\odot}$ being favored  to account for the recorded $\delta$ values \cite{liu14}. Moreover, the distribution of the observed correlated data suggests that these pockets (larger than in most previous parametric attempts) must be associated with a low mass fraction of $^{13}$C to avoid production of $^{14}$N. 

In general, the SiC grain distribution of s-isotopes suggests large, diluted, and almost flat pocket profiles Such characteristics become difficult to be parametrically described, if not with with an ad-hoc and very fine tuning of the parameters. On the contrary, physical models based on MHD-mechanisms for the pocket formation easily succeed now in obtaining a sizable pocket with a quasi-flat distribution of $^{13}$C (see e.g. Fig. \ref{magpocket}), whose size and abundances naturally change according to stellar mass, metallicity and evolutionary stage along the AGB \cite{bu21}. Waiting for dedicated computations in the gravity-wave scenario presented in section \ref{gravity}, we guess that they could in principle obtain similar results. 

\cred{In many cases, the remaining discrepancies between grain abundance ratios and theoretical predictions arise from the difficulty of matching the observed isotopic ratios with carbon-rich stellar models. Although the isotopic distributions of s-process elements (in particular Zr and Ba) can be reproduced relatively easily by the models, a major issue is that the predicted values are reached well before the surface C/O ratio becomes greater than or equal to 1. This problem becomes even more pronounced for stellar models with solar or sub-solar metallicity.
At present, two scenarios have been proposed to  improve the situation. The first assumes that the stellar progenitors of MS-SiC grains are AGB stars with super-solar metallicity  \cite{lug20}. The second retains the classical assumption of solar or sub-solar metallicity progenitors, but invokes additional mixing driven by magnetic transport, which produces C- and s-element–rich regions in the stellar envelope while it is still O-rich  \cite{pal21}. 
Models with higher metallicity experience lower neutron exposures; consequently, the first scenario improves the agreement between observations and predictions, particularly for Sr and Zr abundances, and provides a better fit to the $^{92}$Zr$/^{94}$Zr ratio \cite{lug23}. In the second scenario, the efficiency of enriching the stellar envelope with material processed in the He shell is enhanced for carbon as well as for all s-process elements. Moreover, this efficiency may vary from one star to another, assuming a distribution of magnetic field intensities  as observed by \cite{pascoli} in AGB envelopes.}

\cred{Beyond the stellar physics,} for some specific cases, such as $^{134}$Ba, improving  the nuclear input data may still be crucial. On this purpose, independent studies \cite{taioli}, \cite{li21} highlight that the critical uncertainty associated with $\beta$-decay rates of  $^{134}$Cs in stellar plasmas might possibly affect the resulting per mil shift $\delta\rm{(^{134}{Ba}/^{136}{Ba})}$ computed in AGB models. New theoretical calculations by \cite{taioli} suggest that the temperature dependence of the decay rate of  $^{134}$Cs be significantly less steep than previously assumed, by factors ranging from 2.5 to approximately 30.

\cred{Very recently \cite{ste25} analyzed molybdenum, ruthenium, and barium isotopes in a sample of MS, Y, Z and a few X grains, the data show an overall good agreement with AGB model calculations especially for Mo and Ru, but at the same time the unprecedented precision achieved by the such ultimate measurements allows to record small spreads in the abundances of nuclei most sensitive to the physical conditions under which the s-process occurs ($^{94}$Mo, $^{95}$Mo, $^{100}$Mo, and $^{104}$Ru}) opening the way to further investigations.

\section{Conclusions}
In this review, we have outlined the history and the present status of s-process nucleosynthesis studies, concentrating on the stellar topics that have been among the interests of Roberto Gallino, teacher, colleague, and dear friend of ours, who passed away in 2024 and to whose memory this collection of papers is dedicated. 

There are obviously other connected research lines that could not fit in the limited number of pages of a single article. We must, from this point of view, at least mention that the accurate measurements of neutron capture cross sections, originally one of the main difficulties in affording  the nucleosynthesis processes induced by free neutrons occurring in stars, has in the mean time become a brilliant scientific achievement of our Nuclear Astrophysics community, crucial contributions coming especially  from the modern infrastructure now present at CERN, i.e. n${\_}$TOF \cite{ntof}. To the development of this facility another giant of the field (and another one among the many friends of ours and Roberto’s), Franz K{\"a}ppeler, dedicated the last years of his incredible career.

While we celebrate the memory of these friends and colleagues, we also carry the terrible burden of the premature demise of the first author of this review, Inma Dom\'{\i}nguez, occurred during this same year, after a terrible disease that she faced always with incredible courage and strength. We all pass. Science remains. Its achievements and our small contributions are the best heritage that we, like them, want and invite the others to preserve.

\vspace{0.5cm}

\noindent{\bf Acknowledgments}\\
S.P. acknowledges the PRIN2022 project entitled “$\beta-$DE\-cays and NEutrons captures for astrophysical Branchings (DE\-NEB)" (2022THRKMK). Part of this work was supported by the Spanish project PID2021-123110NB-I00, financed by MICIU/AEI/10.13039/501100011033 and by FEDER, "una manera de hacer Europa", UE. CA dedicates this work to his all life partner Inma Dom\'\i nguez, for her unconditional support, loyalty, and love throughout her life.

\bibliographystyle{elsarticle-num}

\bibliography{c13.bib}
\end{document}